\documentclass{appolb}
\usepackage{epsfig}
 
\begin{document}
\title{Some observational aspects of the orientation of galaxies}
 
\author{W{\l}odzimierz God{\l}owski
\address{Institute of Physics, Opole University, Opole, Poland }
}
 
\maketitle
\begin{abstract}
We investigated the sample of galaxies  belonging to the Tully groups
of galaxies. We analyzed the orientation of galaxies inside the group.
We did not found significant deviation from isotropy both in orientation
of position angles and  angles $\delta_D$ and $\eta$ giving the spatial
orientation of galaxy planes. Moreover we analyzed consequences of different
approximation of "true shape" of galaxies and showed possible influence of
this problem for investigation of spatial orientation of galaxies.
Implications of the obtained results for theory of galaxy formation was 
discussed as well.
\end{abstract}
 
\section{Introduction}
 
The problem of the formation of the structures in the Universe is one of
the most important problem of modern extragalactic astronomy and cosmology.
One of interesting aspects of this problem is the analysis  of the orientation
of galaxies inside the galaxy structures. The very important question is if
there exists  dependence on the alignment to the mass of the analyzed structures.
It is because the investigation of the orientation of galaxies planes is
regarded as a standard test of galaxies formation scenarios.
\cite{Peebles69,Zeldovich70,Sunyaew72,Doroshkevich73,Shandarin74,Dekel85,Wesson82,Silk83,Bower05}.
 
There was a lot of investigation of the orientations of galaxies inside clusters
(see \cite{g11} for the latest review). God{\l}owski et al \cite{g05} suggested
that alignment of galaxies in cluster should increase with the number of
objects in a particular cluster. That suggestion was confirmed by Aryal et al.
\cite{Aryal07}, based on the series of papers \cite{Aryal04,Aryal05,Aryal06}.
However both \cite{g05} and \cite{Aryal07} analysis were qualitative only.
For this manner God{\l}owski at al. \cite{g10a} analyzed sample of 247 rich
Abell clusters using statistical tests and found that alignment increases
with the richness of the clusters.  The analysis of the orientation of galaxies
in poor galaxy structures - Tully groups of galaxies was performed by
God{\l}owski et al. \cite{g05}. In this paper it was found that the group do
not exhibit clear evidence for existence of alignment in the investigated
structures. However they concluded, that observational effect generated by
the process of deprojection of galaxies \cite{g5}, later confirmed by
\cite{g4,Baier03}, masks to the high degree any possible alignment during analysis
of the spatial orientation of galaxies in clusters. For these reasons we
analyzed the orientation of galaxy in Tully galaxy group in more details.

\begin{table}
\begin{center}
\scriptsize
\caption{Test for isotropy of the orientations of galaxy plane. The distribution
of the angle $\delta$ of galaxies, inclination taken directly from NGC Catalog.}
\label{tab:t1}
\begin{tabular}{ccrrrrrcr}
\hline
angle&group&$N$&$\chi^2$&$C$&$P(\Delta_1)$&$\Delta_{11}$&$\sigma(\Delta_{11})$&$\lambda$\\
\hline
        & $11$ & 626&  62.8&   9.50& .000& -.237& .058& 1.20\\
        & $12$ & 332&  25.7& -11.48& .299& -.125& .080& 0.55\\
        & $13$ & 128&  29.7&  -6.94& .891& 0.002& .129& 0.62\\
        & $14$ & 426&  24.0&   3.52& .154& -.120& .071& 0.78\\
        & $15$ & 130&  13.1&  -1.88& .737& -.004& .128& 0.59\\
        & $17$ &  80&  13.8&  -5.50& .569& -.100& .164& 0.82\\
        & $21$ & 248&  14.2&   0.24& .065& -.035& .093& 1.14\\
        & $22$ & 126&   7.0&   0.68& .496& 0.098& .130& 0.62\\
$\delta$& $23$ & 100&  17.0&  -1.82& .607& 0.140& .146& 0.47\\
        & $31$ & 210&  33.7&  15.98& .000& 0.137& .101& 1.79\\
        & $41$ & 192&  22.8&   6.54& .020& -.004& .106& 1.37\\
        & $42$ & 230&  24.7&  -3.78& .320& 0.056& .096& 0.75\\
        & $44$ &  80&  26.7&   7.95& .024& -.200& .164& 1.34\\
        & $51$ & 228&  29.6&  -0.41& .009& 0.093& .097& 1.44\\
        & $52$ & 172&  21.6&   3.04& .005& -.203& .112& 1.42\\
        & $53$ & 260&  13.3&  -3.29& .492& -.055& .091& 0.62\\
        & $61$ & 258&  19.9&  -3.13& .825& -.049& .091& 0.62\\
        & $64$ & 102&  28.7&  -7.54& .113& 0.080& .145& 0.84\\
\hline
\end{tabular}
\end{center}
\end{table}

\begin{table}
\begin{center}
\scriptsize
\caption{Test for isotropy of the orientations of galaxy plane. The distribution
of the angle $\eta$ of galaxies, inclination taken directly from NGC Catalog.}
\label{tab:t2}
\begin{tabular}{ccrrrrrcr}
\hline
angle&group&$N$&$\chi^2$&$C$&$P(\Delta_1)$&$\Delta_{11}$&$\sigma(\Delta_{11})$&$\lambda$\\
\hline
        & $11$ & 626&  60.0&   5.96& .000& 0.304& .057& 2.00\\
        & $12$ & 332&  28.5&   7.56& .001& -.069& .078& 1.70\\
        & $13$ & 128&  25.6&   2.78& .079& 0.242& .125& 0.77\\
        & $14$ & 426&  27.4&   6.63& .090& 0.055& .069& 1.00\\
        & $15$ & 130&  22.9&   2.51& .764& 0.036& .124& 0.91\\
        & $17$ &  80&  13.1&  -5.97& .470& -.059& .158& 0.46\\
        & $21$ & 248&  26.9&  -3.55& .054& 0.058& .090& 1.23\\
        & $22$ & 126&  11.7&   5.57& .177& 0.194& .126& 0.98\\
$\eta$  & $23$ & 100&  20.2&  -0.10& .081& -.164& .141& 1.09\\
        & $31$ & 210&  24.0&   0.43& .046& 0.194& .098& 1.56\\
        & $41$ & 192&  27.2&  10.22& .001& 0.300& .102& 1.95\\
        & $42$ & 230&  15.9&   3.30& .033& 0.036& .093& 1.24\\
        & $44$ &  80&  20.4&  -3.05& .226& 0.148& .158& 0.78\\
        & $51$ & 228&  30.6&  -5.37& .042& 0.218& .094& 1.15\\
        & $52$ & 172&  38.1&  12.29& .001& 0.212& .108& 1.92\\
        & $53$ & 260&  12.2&  -8.28& .816& 0.008& .088& 0.34\\
        & $61$ & 258&  23.7& -12.56& .549& 0.091& .088& 0.64\\
        & $64$ & 102&  50.1&  -3.53& .002& 0.402& .140& 1.88\\
\hline
\end{tabular}
\end{center}
\end{table}

\begin{table}
\begin{center}
\scriptsize
\caption{Test for isotropy of the distribution of supergalactic position
angles $P$ of galaxies.}
\label{tab:t3}
\begin{tabular}{ccrrrrrcr}
\hline
angle&group&$N$&$\chi^2$&$C$&$P(\Delta_1)$&$\Delta_{11}$&$\sigma(\Delta_{11})$&$\lambda$\\
\hline
    & $11$ &185 & 22.7& -11.42& .728&  0.081& .104& 0.53\\
    & $12$ &106 & 17.6&  -1.57& .198&  -.083& .137& 0.88\\
    & $13$ & 50 & 13.4&  -2.48& .714&  -.056& .200& 0.41\\
    & $14$ &133 & 14.9&  -1.05& .990&  -.011& .123& 0.49\\
    & $15$ & 48 & 11.3&  -0.75& .185&  -.006& .204& 1.06\\
    & $17$ & 22 & 10.7&  -5.64& .727&  -.152& .302& 0.62\\
    & $21$ & 85 & 13.5&  -2.84& .878&  -.058& .153& 0.67\\
    & $22$ & 43 & 16.0&  -1.98& .910&  -.089& .216& 0.63\\
$P$ & $23$ & 33 & 12.3&   0.27& .230&  -.337& .246& 0.81\\
    & $31$ & 63 & 20.1&   1.00& .729&  0.081& .178& 0.82\\
    & $41$ & 54 & 20.0&  10.67& .595&  -.112& .192& 1.09\\
    & $42$ & 71 & 19.0&   1.51& .124&  -.254& .168& 0.91\\
    & $44$ & 25 & 18.9&   1.64& .367&  0.161& .283& 0.98\\
    & $51$ & 69 & 23.1&   3.00& .576&  -.176& .170& 0.88\\
    & $52$ & 50 & 14.8&  -0.32& .631&  0.154& .200& 0.77\\
    & $53$ & 88 & 17.5&   2.82& .080&  0.243& .151& 1.36\\
    & $61$ & 85 & 27.9&   6.48& .007&  -.116& .153& 1.68\\
    & $64$ & 31 & 18.4&   2.10& .295&  -.397& .254& 0.91\\
\hline
\end{tabular}
\end{center}
\end{table}
 
\section{Observational data}
 
The aim of our work is to study the alignment of galaxies in galaxy groups.
Groups were taken from Tully Nearby Galaxies (NBG) Catalog \cite{t3}. This 
Catalog contains 2367 galaxies with radial velocities less than $3000\,km\,s^{-1}$.
Tully Catalog provides relatively uniform coverage of entire unobscured
sky \cite{t2}. Galaxies position angles were taken from \cite{n1,n2,l1,l2}
while some missing measurements  were made on PSS prints by Piotr Flin
\cite{g10}. The NBG Catalog gives the group affiliation for the galaxies
belonging to the catalog. The groups extracted from the NBG Catalog
are one of the best selections with precise criterion of groups membership.
Moreover, the  galaxy distances are very  well and in uniform maner
determined. As a result the lists of galaxies belonging to the particular
groups are free from the background objects which is crucial in such type 
of the analysis. From the NBG Catalog we extracted structures having
at least 40 members.

\begin{table}[h]
\begin{center}
\scriptsize
\caption{The results of numerical simulations for positions angles $P$}.
\label{tab:t4}
\begin{tabular}{ccccc}
\hline
Test&$\bar{x}$&$\sigma(x)$&$\sigma(\bar{x})$&$\sigma(\sigma(x))$\\
\hline
  $\chi^2$                       & 16.9524&  1.4592&   0.0461&  0.0326 \\
  $\Delta_{1}/\sigma(\Delta_{1})$&  1.2513&  0.1543&   0.0048&  0.0034 \\
  $\Delta/\sigma(\Delta)$        &  1.8772&  0.1581&   0.0050&  0.0035 \\
\hline
\end{tabular}
\end{center}
\end{table}

\begin{table}[h]
\begin{center}
\scriptsize
\caption{The statistics of the observed distributions for real clusters}
\label{tab:t5}
\begin{tabular}{cc|cc|cc|cc}
\hline
\multicolumn{2}{c}{}&
\multicolumn{2}{c}{$P$}&
\multicolumn{2}{c}{$\delta_D$}&
\multicolumn{2}{c}{$\eta$}\\
\hline
Sample&Test&$\bar{x}$&$\sigma(x)$&$\bar{x}$&$\sigma(x)$&$\bar{x}$&$\sigma(x)$\\
\hline
 &$\chi^2$                       & 17.338& 1.061& 23.794& 2.853& 26.583& 2.946\\
A&$\Delta_{1}/\sigma(\Delta_{1})$&  1.282& 0.177&  1.847& 0.257&  2.443& 0.287\\
 &$\Delta/\sigma(\Delta)$        &  2.112& 0.209&  2.328& 0.256&  2.910& 0.272\\
\hline
 &$\chi^2$                       & 16.800& 1.152& 18.700& 1.332& 20.188& 2.166\\
B&$\Delta_{1}/\sigma(\Delta_{1})$&  1.218& 0.176&  1.420& 0.169&  1.554& 0.233\\
 &$\Delta/\sigma(\Delta)$        &  2.081& 0.218&  2.076& 0.165&  2.385& 0.302\\
\hline
\end{tabular}
\end{center}
\end{table}

\begin{figure}
\begin{center}$
\begin{array}{ll}
\includegraphics[scale=0.95]{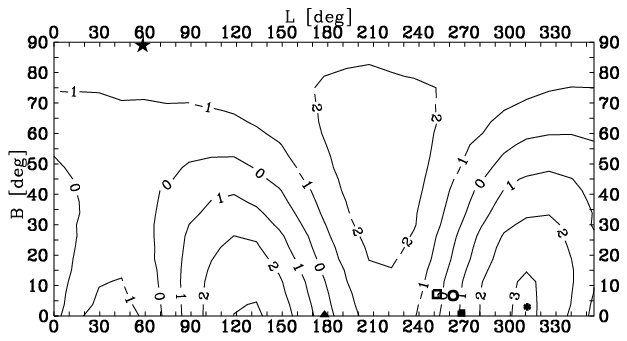} & \includegraphics[scale=0.95]{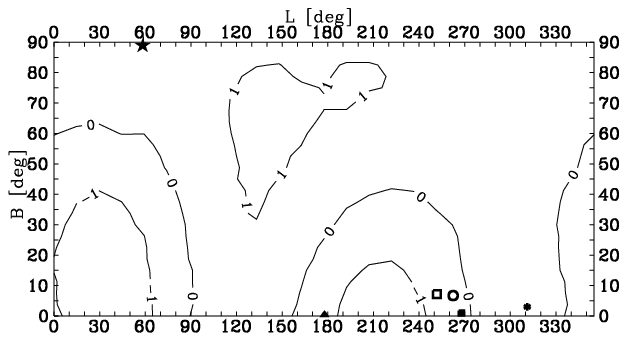}\\
\includegraphics[scale=0.95]{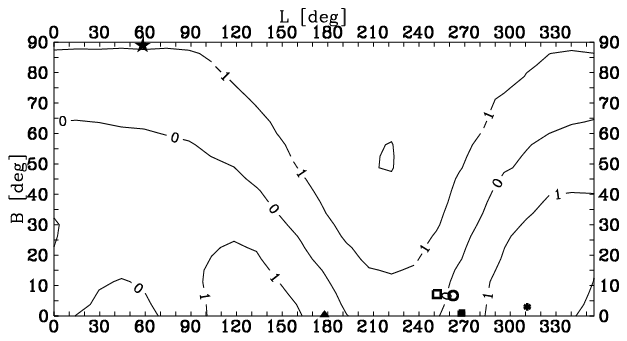} & \includegraphics[scale=0.95]{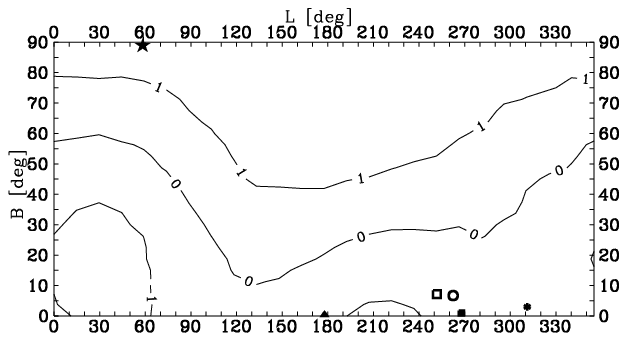}\\
\includegraphics[scale=0.95]{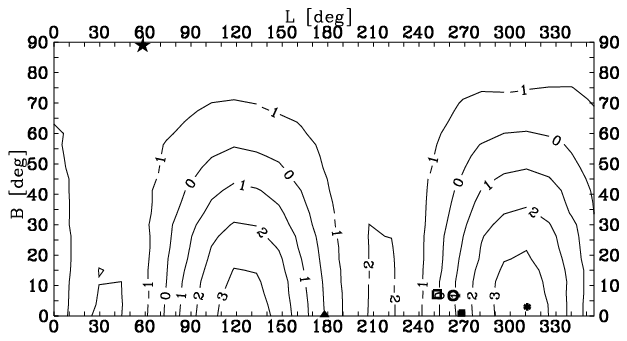} & \includegraphics[scale=0.95]{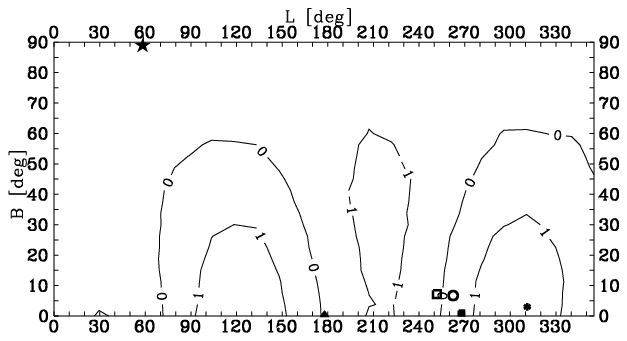}\\
\end{array}$
\end{center}
\scriptsize
\caption{
Maps of $s \equiv \Delta_{11} / \sigma(\Delta_{11})$ versus the
chosen cluster pole supergalactic co-ordinates ($L$, $B$) for the cluster 12.
In the maps, the results of the Tully data are shown on the left, while result
obtained with help of HHD and FP corrections are given
on the right.  The maps are presented for ALL cluster
galaxies (upper panel), for S (middle panel) and NS (bottom panel)
sub-samples. In the map we indicate the
important directions, as seen from the centre of the considered cluster:
1.) three cluster poles (full star, square and triangle), 2.)
the direction to the Local Supercluster centre (open circle), 3.) the
direction of the Virgo A cluster centre (open square) and 4.) the line
of sight from the Earth (asterisk).}
\end{figure}
 
\section{Methods of the analysis and results}
 
We studied the alignment of galaxies in Tully groups of galaxies belonging
to the Local Supercluster (LSC).  Till now two main methods for study of the
galaxy orientation were proposed. In the first one \cite{h4}
the distribution of the position angle of the galactic image major 
axis was analyzed. In this approach, face-on and nearly face-on
galaxies were excluded from the analysis and  only  galaxies with
axial ratio $d/D \le 0.75$  were taken into consideration.
The second approach, based on the de-projection of the galaxy images,
allowed us to use also the face-on galaxies, This method was originally
proposed by {\"O}pik \cite{Op70},  applied by Jaaniste \& Sarr \cite{Ja78}
and significantly modified by Flin \& God{\l}owski \cite{f4,f5,f6,g2,g3,g10}
In this method  the  galaxy's inclination with respect to the observer's
line of sight $i$ is  considered.
In the Tully NBG Catalog \cite{t3} the inclination angle was calculated
according  to formula  $i=cos^{-1}{(q^2 -q^2_0 )/(1-q^2_0)}^{-1/2}+3^0$,
where $q=d/D$ is the ratio of minor to major axis diameters and
$q_0$  is  "true" axial ratio.  Tully used standard value $q_0=0.2$.
One should note that above formula is the modified Holmberg's \cite{Holmberg46}
formula valid for oblate spheroids. For each galaxy, two angles are
determined: $\delta_D$ - the angle between the normal to the galaxy plane
and the main plane of the coordinate system, and $\eta$ - the angle between
the projection of this normal onto the main plane and the direction towards
the zero initial meridian. Using the Supergalactic coordinate system
(Flin \& God{\l}owski \cite{f4} based on \cite{TS76}) the following relations
hold between angles ($L$, $B$, $P$) and ($\delta_D$, $\eta$)
\begin{equation}
\sin\delta_D  =  -\cos{i}\sin{B} \pm \sin{i}\cos{r}\cos{B},
\end{equation}
\begin{equation}
\sin\eta  =  (\cos\delta_D)^{-1}[-\cos{i}\cos{B}\sin{L} + \sin{i}
(\mp \cos{r}\sin{B}\sin{L} \pm \sin{r}\cos{L})],
\end{equation}
where $r=P-\pi/2$.
 
In order to detect non-random effects in the distribution of the
investigated angles: $\delta_D$, $\eta$ and $P$ we divided the entire
range of the analyzed angles into $18$ bins and carried
out three different statistical tests. These tests were : the $\chi^2$ test,
the autocorrelation test and the Fourier test  \cite{h4,g2,g3,g10a,g11a}. For
$n=18$ the $\chi^2$ test yields critical value $27.59$ (at the significance
level $\alpha=0.05$) while critical value for autocorrelation test is
$C_{cr}\approx 6.89$. The last value is obtained from numerical simulations
using the method described by God{\l}owski \cite{g11a}. The isotropy of the
resultant distributions of the investigated angles was also analyzed using
Kolmogorov-Smirnov test (K-S test). We assumed that the theoretical, random
distribution contains the same number of objects as the observed one. In order
to reject the $H_0$ hypothesis, that the distribution is random one, the value
of observed statistics $\lambda$ should be greater than $\lambda_{cr}= 1.358$
(for $\alpha=0.05$). One should note however that, especially in the case of
position angles, the number of analyzed galaxies is sometimes small and does not
satisfy theoretical tests conditions. It is the reason that we repeated our
analysis with different numbers of bins, founding insignificant differences
in these cases.

\begin{table}
\begin{center}
\scriptsize
\caption{Test for isotropy of the orientations of galaxy plane.
The distribution of the angle $\delta$ of galaxies, inclination obtained
according to HHD and FP corrections.}
\label{tab:t6}
\begin{tabular}{ccrrrrrrcr}
\hline
angle&group&$N$&$\chi^2$&$C$&$P(\Delta_1)$&$\Delta_{11}$&$\sigma(\Delta_{11})$&$\lambda$\\
\hline
        & $11$ & 626&  14.3&  -2.46& .504& -.068& .058& 0.39 \\
        & $12$ & 332&  12.4&   2.49& .744& 0.059& .080& 0.37 \\
        & $13$ & 128&   8.7&   0.86& .387& 0.161& .129& 0.54 \\
        & $14$ & 426&  22.7&   1.50& .654& -.063& .071& 0.52 \\
        & $15$ & 130&  26.2&   4.40& .025& 0.243& .128& 1.25 \\
        & $17$ &  80&   7.0&  -1.57& .949& -.018& .164& 0.48 \\
        & $21$ & 248&  19.7&   6.05& .575& 0.067& .093& 0.85 \\
        & $22$ & 126&  17.9&   1.77& .037& 0.187& .130& 1.30 \\
$\delta$& $23$ & 100&  22.1&  -3.00& .150& 0.275& .146& 0.71 \\
        & $31$ & 210&  19.1&   3.90& .049& 0.050& .101& 0.74 \\
        & $41$ & 192&  20.9&  -4.77& .354& 0.134& .106& 0.69 \\
        & $42$ & 230&  18.3&  -4.08& .419& -.091& .096& 0.86 \\
        & $44$ &  80&  23.9&   0.74& .525& -.104& .164& 0.93 \\
        & $51$ & 228&  18.3&   3.40& .793& 0.031& .097& 0.61 \\
        & $52$ & 172&  20.7&   4.97& .005& -.169& .112& 1.53 \\
        & $53$ & 260&  13.2&  -1.46& .669& -.061& .091& 0.62 \\
        & $61$ & 258&  25.3&  -1.26& .359& -.020& .091& 1.03 \\
        & $64$ & 102&  25.9&  -1.99& .146& 0.008& .145& 0.78 \\
\hline
\end{tabular}
\end{center}
\end{table}

\begin{table}
\begin{center}
\scriptsize
\caption{Test for isotropy of the orientations of galaxy plane.
The distribution of the angle $\eta$ of galaxies,  inclination
obtained according to HHD and FP corrections. }
\label{tab:t7}
\begin{tabular}{ccrrrrrcr}
\hline
angle&group&$N$&$\chi^2$&$C$&$P(\Delta_1)$&$\Delta_{11}$&$\sigma(\Delta_{11})$&$\lambda$\\
\hline
        & $11$ & 626&  44.1&  13.00& .000& 0.215& .057& 1.37\\
        & $12$ & 332&  21.3&   3.87& .075& -.070& .078& 1.65\\
        & $13$ & 128&  18.3&   5.17& .272& 0.191& .125& 0.67\\
        & $14$ & 426&   8.5&   0.34& .418& 0.038& .069& 0.45\\
        & $15$ & 130&  22.0& -11.48& .651& 0.089& .124& 0.57\\
        & $17$ &  80&   8.2&  -3.05& .603& -.103& .158& 0.40\\
        & $21$ & 248&  15.8&  -6.45& .718& 0.015& .090& 0.75\\
        & $22$ & 126&  14.6&   1.00& .767& 0.092& .126& 0.53\\
$\eta$  & $23$ & 100&   7.3&  -1.18& .809& 0.009& .141& 0.40\\
        & $31$ & 210&  20.2&   0.17& .047& 0.226& .098& 0.81\\
        & $41$ & 192&  38.3&  20.16& .007& 0.099& .102& 1.61\\
        & $42$ & 230&  20.6&   4.94& .729& 0.074& .093& 0.59\\
        & $44$ &  80&  19.5&  -3.95& .763& -.013& .158& 0.55\\
        & $51$ & 228&  20.8&   2.61& .736& 0.067& .094& 0.66\\
        & $52$ & 172&  17.6&  -0.37& .085& 0.100& .108& 0.99\\
        & $53$ & 260&  19.3&  -3.50& .888& -.005& .088& 0.51\\
        & $61$ & 258&  23.7&   2.09& .045& 0.122& .088& 1.22\\
        & $64$ & 102&  23.3&  -5.65& .289& 0.204& .140& 0.76\\
\hline
\end{tabular}
\end{center}
\end{table}

\begin{table}
\begin{center}
\scriptsize
\caption{Test for isotropy of the distribution of supergalactic position
angles $P$ of galaxies. Only galaxies with certain measure $P$ are taken
into account.}
\label{tab:t8}
\begin{tabular}{ccrrrrrcr}
\hline
angle&group&$N$&$\chi^2$&$C$&$P(\Delta_1)$&$\Delta_{11}$&$\sigma(\Delta_{11})$&$\lambda$\\
\hline
    & $11$ & 143& 17.7& -9.32& .991& -.001& .118& 0.40\\
    & $12$ &  96& 19.5&  0.19& .113& -.071& .144& 0.99\\
    & $13$ &  40& 16.7&  1.40& .805& -.142& .224& 0.81\\
    & $14$ & 114& 14.2&  2.21& .943& -.045& .132& 0.44\\
    & $15$ &  42& 12.0& -0.86& .201& 0.057& .218& 1.03\\
    & $17$ &  21& 10.7& -5.57& .656& -.241& .309& 0.58\\
    & $21$ &  76& 11.2& -3.53& .859& -.059& .162& 0.55\\
    & $22$ &  37& 15.1& -1.49& .997& -.003& .232& 0.57\\
$P$ & $23$ &  33& 12.3&  0.27& .230& -.337& .246& 0.81\\
    & $31$ &  58& 23.3&  4.07& .444& 0.080& .186& 0.83\\
    & $41$ &  46& 21.3&  9.17& .481& -.121& .209& 1.08\\
    & $42$ &  64& 18.1&  2.38& .075& -.282& .177& 1.04\\
    & $44$ &  21& 19.3& -2.14& .352& 0.245& .309& 0.84\\
    & $51$ &  60& 21.0&  0.00& .759& -.110& .183& 0.82\\
    & $52$ &  32&  8.5& -1.63& .788& 0.170& .250& 0.37\\
    & $53$ &  84& 16.3&  0.86& .135& 0.203& .154& 1.24\\
    & $61$ &  78& 27.7&  1.38& .040& -.086& .160& 1.47\\
    & $64$ &  25& 17.5&  3.08& .331& -.362& .283& 1.14\\
\hline
\end{tabular}
\end{center}
\end{table}

At first, following God{\l}owski et al. \cite{g05} but adding the K-S test,
we analyzed orientation of galaxies in Tully groups using, for obtaining
$\delta_D$ and $\eta$ angles, inclinations angles taken directly from NBG
Catalog \cite{t3} (sample A). The results were presented in the Tables 1-3.
Analysis of the supergalactic position angles showed that only one group 
($61$) exhibits alignment of galaxies. Analysis of the distribution of the 
angles  giving spatial orientation of galaxies ($\delta_D$ and $\eta$) seems 
to show weak alignment. For $\delta_D$ angle three tests showed that the 
distribution is non random in the case of the clusters $11$, $31$ and $51$. 
Two tests showed nonrandomness in the case of clusters $41$ and $52$. For 
$\eta$ angle three tests showed alignment in the case of clusters $11$, $12$, 
$41$, $52$ and $64$. Two tests showed it in the case of clusters $31$ and $51$.
  
For more detailed analysis we used the method described in \cite{g11a}.
The question which arose,  is if we could say that we found an alignment
in the analyzed sample of $18$ Tully groups of galaxies. So we computed the
mean value  and variance of analyzed statistics:
$\chi^2$, $\Delta_1/\sigma(\Delta_1)$, $\Delta/\sigma(\Delta)$
(i.e. the same statistics as was analyzed in \cite{g10a}) for our sample
of $18$ groups and compared it with results of numerical simulations. We performed
1000 simulations of 18 fictious clusters, each with number of randomly oriented,
members galaxies, the same as in real clusters. In the Table 4
we present, obtained from numerical simulations average values of the analyzed
statistics, their standard deviations, standard deviations in the sample as well
as their standard deviations for distribution of $P$ angles. One should note that
there are some differences in results of numerical simulations  for $P$,
$\delta_D$ and $\eta$ angles but it does not change our further conclusions. The
mean values and variance of analyzed statistics  for sample of real clusters
are presented in the Table 5. On can show that (for sample A) analysis of
the position angles does not show significant deviation from the values expected
in the case of random distributions, while it seems that analysis of angles
$\delta_D$ and $\eta$ shows existence of alignment at the $2\sigma$ level
(with exception of $\Delta/\sigma(\Delta)$ statistics for $\delta_D$ angle). 
However, below we will argue against such interpretation.
 
The Tully groups were analyzed also by God{\l}owski and Ostrowski \cite{g5}.
For every cluster the parameter $\Delta_{11}$
describing  the galactic axes alignment with respect  to  a chosen
cluster pole, divided by its formal error $\sigma(\Delta_{11})$ ($s
\equiv \Delta_{11} / \sigma(\Delta_{11})$) were mapped. The cluster pole
coordinates change  along the entire  celestial sphere.  The  resulting
maps  were  analyzed  for correlations  of their maxima with important
points on the maps (see Fig.1 for details). It was found that maxima
correlated well with the line of  sight  direction.
God{\l}owski and Ostrowski \cite{g5}  concluded that
the strong systematic effect, generated by the process of galactic axis
de-projection from its optical image, is present in the catalogue data.
The example of that effect is presented in the Fig.1. To avoid possible
influence of global alignment inside LS, we choose for presentation
the cluster $12$ (Ursa Major Cloud) because it is far from the center
of the Local Supercluster (i.e. Virgo Cluster).
To remove above effect we should avoid the assumption that the "true"
axial ratio is $q_0=0.2$, which is a rather poor approximation, especially for 
non-spiral galaxies. Fortunately NBG Catalog contains morphological types of 
galaxies. These allowed us to use different values of $q_0$ depending on morphological
type \cite{hhv}. Now, with help of Fouque \& Paturel \cite{fp85} formulae, which
convert $q$ to standard photometrical axial ratios, we compute new inclination
angle $i$ for all galaxies in NGB catalog. We repeated our investigation
with that "new" sample of galaxies (sample B). As one can see from right
panel of the Fig.1, the "line of sight" effect disappeared which shows 
that our procedure for computation of the inclination angles are much 
better than the previous one.
From the Tab.6 and Tab.7 we can show that during analysis of the spatial
orientation of galaxies, the alignment is observed only in the case of $\eta$ angle
for clusters $11$ and $41$ (Virgo Cluster and Virgo - Libra Cloud).
Of course our procedure does not change position angles. In the Tab.8 we presented
analyzes of the distribution of $P$ where only galaxies with the
certainly measure positions angles were taken into account. On can show
(Tab.5, sample $B$)  that now the mean value of analyzed statistics
does not exhibit any significant deviation from the values expected
in the case of random distributions. The above results allowed us to conclude
that we did not observe any significant alignment for Tully groups of galaxies.

\section{Discussion and conclusions}

We investigated the orientation of galaxies inside $18$ Tully groups of galaxies
belonging to the Local Supercluster, do not founding any significant alignment.
So we conclude that orientations of galaxies in the Tully groups are random.
We also analyzed observational effect generated by the process of 
deprojection of  galaxies found by God{\l}owski and Ostrowski \cite{g5},
which masks to the high degree any possible alignment during analysis of the
spatial orientation of galaxies in clusters. We showed that using "true shape"
of galaxies $q_0$ depending on morphological type according to Heidmann et al.
\cite{hhv} with help of Fouque \& Paturel \cite{fp85} corrections of $q$
to standard photometrical axial ratios, allowed us to avoid this problem. This 
gives much more powerful investigation of the spatial orientation of galaxies.
 
In our prvious papers we found that in the sample of $247$ rich Abell galaxy
clusters we observed an alignment, which increased with the cluster richness
\cite{g10a,g11a}. Now, we found that the orientation of galaxies inside the
poor galaxy structure is random. It confirms our suggestion that alignment
of galaxies increases with the mass of the structures \cite{g11}.
Usually such dependence between the angular momentum and the mass of the
structure is presented as empirical relation $J\sim M^{5/3}$
\cite{Wesson79,Wesson83,Carrasco82,Brosche86}.  The observed relation between
mass of the galaxy structure and the alignment is compatible with the prediction
of the Li model \cite{Li98} in which galaxies are forming in the rotating universe.
However, in our opinion, it is due to tidal torque, as suggested by
\cite{HP88,Catelan96}. Also the results of the analysis of the linear
tidal torque theory which  noticed the connection of the alignment with the
considered scale of the structure is pointing in the same direction
\cite{Noh06a,Noh06b}.

\end{document}